\documentclass{ar-1col-S2O}
\usepackage[numbers]{natbib}
\usepackage{url}
\usepackage{bm}
\usepackage{braket}
\newcommand{\etal}{\textit{et al}. }
\newcommand{\ie}{\textit{i}.\textit{e}., }
\newcommand{\eg}{\textit{e}.\textit{g}.\ }
\usepackage[colorlinks,linkcolor=blue,citecolor=blue,urlcolor=blue]{hyperref}
\usepackage{tikz}

\begin{document}

% Page header
\markboth{Yan}{Chirality}

% Title
%\title{Interplay of Chirality, Spin and Orbital in Chiral Quantum Materials}
\title{Structural Chirality and Electronic Chirality in Quantum Materials}

%Authors, affiliations address.
\author{Binghai Yan
\affil{Department of Condensed Matter Physics, Weizmann Institute of Science, Rehovot 7610001, Israel}
}
%Abstract
\begin{abstract}
%Abstract text, approximately 150 words. 
In chemistry and biochemistry, chirality represents the structural asymmetry characterized by non-superimposable mirror images for a material like DNA. In physics, however, chirality commonly refers to the spin-momentum locking of a particle or quasiparticle in the momentum space. 
While seemingly disconnected, structural chirality in molecules and crystals can drive electronic chirality through orbital-momentum locking, i.e. chirality can be transferred from the atomic geometry to electronic orbitals.
 %While seemingly unrelated characters in different fields, the structural chirality leads to the electronic chirality featured by the orbital-momentum locking encoded in the wavefunction of chiral molecules or solids, \ie the chirality information transfers from the atomic geometry to the electronic orbital. 
 Electronic chirality provides an insightful understanding of the chirality-induced spin selectivity (CISS), in which electrons exhibit salient spin polarization after going through a chiral material, and electric magnetochiral anisotropy (EMCA), which is characterized by the diode-like transport. It further gives rise to new phenomena, such as anomalous circularly polarized light emission (ACPLE), in which the light handedness relies on the emission direction. These chirality-driven effects will generate broad impacts in fundamental science and technology applications in spintronics, optoelectronics, and biochemistry. 
\end{abstract}

%Keywords, etc.
\begin{keywords}
%keywords, separated by comma, no full stop, lowercase
handedness, topology, orbital, spin, orbital-momentum locking, Berry phase, spin valve, circularly polarized light, Onsager's relation
\end{keywords}
\maketitle

%Table of Contents
\tableofcontents

% Heading 1
\section{INTRODUCTION}
\begin{figure}
    \centering
    \includegraphics[width=0.8\textwidth]{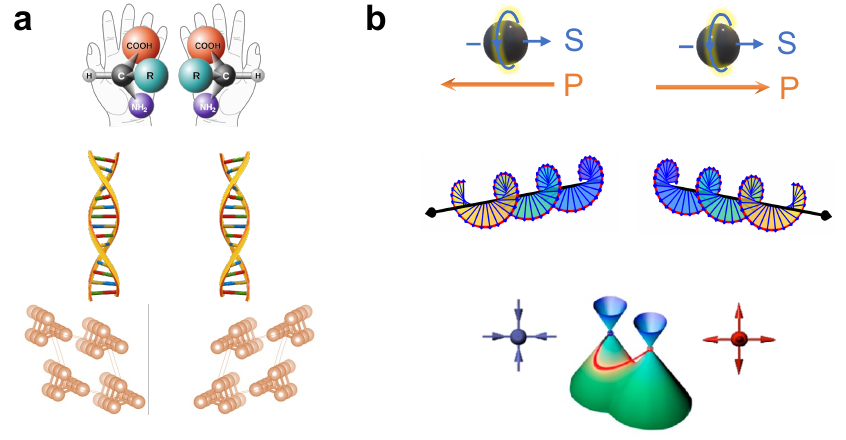}
    \caption{Examples of chirality in real space and in momentum space. (a) Chiral images (enantiomers) for chiral molecules, DNA and chiral crystals regarding the atomic structure. (b) Chiral images for neutrinos, photons (circular polarization), and Weyl fermions regarding the parallel/anti-parallel spin-momentum locking. }
    \label{fig:chirality}
\end{figure}

Chirality represents the geometric nature of a diverse array of molecules (such as DNA or sugar) characterized by non-superimposable mirror images, commonly referred to as left- or right-handed forms (see \textbf{Figure \ref{fig:chirality}a}). This concept has remained a fundamental cornerstone in the realms of both chemistry and biology for more than one century ~\cite{kelvin1894molecular,Siegel1998}. Notably, it has also attracted increasing interest within the domains of physics and materials science in recent years. In physics, chirality usually refers to the locking of spin and momentum in the reciprocal space. For example, the Weyl fermion \cite{Yan2017,Armitage2017}, neutrinos, or circularly polarized light (CPL) is identified to be right(left)-handed if its spin aligns parallel (anti-parallel) to the momentum(\textbf{Figure ~\ref{fig:chirality}b}). 

Although chirality represents seemingly unrelated characters in different fields, recent interdisciplinary experiments~\cite{Gohler2011,Xie2011}
unveiled a compelling connection between the chiral atomic structure and electron spin polarization.
When transmitted through chiral molecules, electrons become spin-polarized, and the spin polarization depends on the material chirality. This effect is called chirality-induced spin selectivity (CISS)~\cite{Naaman2012,Naaman2019}, which can generate larger spin polarization than an ordinary ferromagnet in an unprecedented manner.
The CISS effect promises immense potential for a wide range of applications such as novel spintronic devices \cite{Dor2013,Naaman2015spintronics,Dor2017magnetization,Yang2021chiralspintronics}, chiral electrocatalysis \cite{Mtangi2015,Liang2022,vadakkayil2023chiral}, and enantiomer selectivity \cite{BanerjeeGhosh2018}. It holds far-reaching fundamental implications involving the interplay of structural chirality, electron spin, and topological orbital. However, the microscopic mechanism of CISS is still actively explored \cite{Evers2020RMP,evers2021theory}. 

In this review, we start with a basic question -- What is the fundamental role of chirality in shaping the electronic states within a chiral material? We will show pedagogically that structural chirality leads to the intrinsic electronic chirality, characterized by the orbital-momentum locking in the quantum wavefunction \cite{liu2021chirality}. 
Then the orbital-momentum locking leads to a spin polarizer/filter assisted by the spin-orbit coupling (SOC) from metal contacts ~\cite{liu2021chirality,Gersten2013,Xiong2023interplay}, giving rise to the CISS effect in chiral molecules and also explaining the electric magnetochiral anisotropy (EMCA) \cite{Rikken2001}, another chirality-driven MR that exists in chiral conductors. We will compare CISS and EMCA and discuss the remaining challenges to understanding MR caused by CISS. 
Except for CISS and EMCA, the orbital-momentum locking also leads to an exotic CPL emission from organic light-emitting diodes (OLEDs) constructed from chiral organic semiconductors, which roots in the topological Berry phase in the optical transition~\cite{wan2023anomalous}.

%Heading 1
\section{ELECTRONIC CHIRALITY}\label{sec:locking}
\subsection{A Toy Model of Chiral Wavefunction}
\begin{figure}
    \centering
    \includegraphics[width=\textwidth]{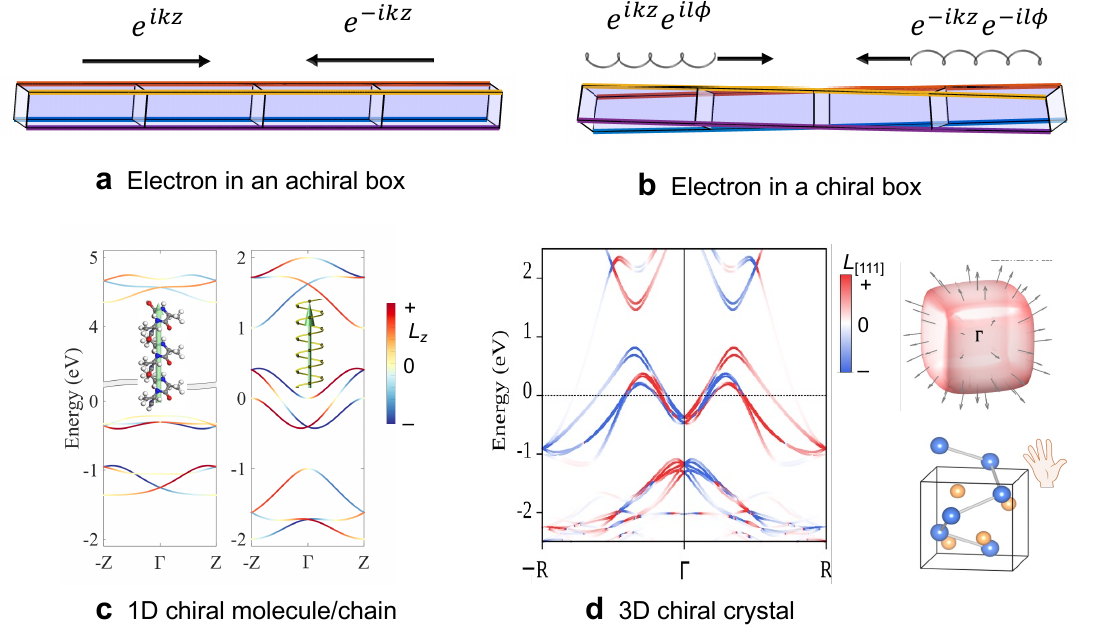}
    \caption{The electronic chirality, \ie orbital-momentum locking in the electronic wavefunction, due to structural chirality. (a) The wavefunction of the electron confined in an ordinary (achiral) box can be regarded as a superposition of counter-propagating plane waves, $e^{\pm ±ikz}$. (b) Similarly, wavefunction in a chiral box is a superposition of two chiral waves, $e^{ikz}e^{il\phi}$ and $e^{-ikz}e^{-il\phi}$, which carry opposite orbital angular momentum ($\pm l$) and share the same electronic chirality ($\chi_e=\mathrm{sign}(\mathbf{k\cdot l})$) induced by the box chirality. The black arrows represent the propagating momentum ($\mathbf{k}$), and the spiral trajectories indicate $\mathbf{l}$. (c) Band structures of a chiral molecule chain ($3_{10}$ helix) from density-functional theory (DFT) (left) and a tight-binding chiral chain (right). The color bar represents the orbital angular momentum ($L_z$) along the chain direction. (d) Band structure of a 3D chiral crystal PdGa from DFT. A Fermi surface near $\Gamma$ exhibits a monopole-like orbital texture (right), \ie orbital-momentum parallel locking along all directions. Adopted from Refs.~\cite{wan2023anomalous,liu2021chirality,yang2023monopole}.}
    \label{fig:orbital}
\end{figure}

We first investigate the electronic wavefunction within a chiral molecule in a heuristic way. 
If we approximate an ordinary (achiral) molecule as a 3D box confining an electron inside, then we treat a chiral molecule as a ``twisted'' 3D box along its long axis ($z$), as illustrated in \textbf{Figure~\ref{fig:orbital}}. The twisting direction (clockwise or counter-clockwise) leads to chirality. Without loss of generality, we take the ground-state wavefunction as an example. Recalling textbook exercises, the ground-state wavefunction  for an ordinary box is 
\begin{equation} \label{eq:planewave}
    \psi = f(\mathbf{r}) \mathrm{cos}(kz) = f(\mathbf{r})(e^{ikz}+e^{-ikz})/2
\end{equation}
where $k=\pi/c$, $c$ is the box length along $z$ and we include other coefficients in $f(\mathbf{r})$ for simple. The $e^{\pm ikz}$ reminds us that the wavefunction (a standing wave solution)  is a superposition of two counter-propagating plane waves, i.e., the electron is moving back-and-forth with a finite momentum ($\pm k$) inside the box (as illustrated in \textbf{Figure~\ref{fig:orbital}a}). If we twist the box slightly in a right-handed way, the original plane wave $e^{ikz}$ ($e^{-ikz}$) will exhibit counter-clockwise (clockwise) rotation in the $xy$ plane when propagating along $+z$ ($-z$) because it feels the boundary rotation. The counter-clockwise (clockwise) rotation means an intrinsic orbital angular momentum (OAM) $+l$ ($-l$). So, we can approximately derive the ``twisted'' wavefunction as
\begin{equation} \label{eq:twistwave}
    \psi = \sum_l g(\mathbf{r})(e^{ikz}e^{il\phi}+e^{-ikz}e^{-il\phi})/2
\end{equation}
where $\phi$ is the cylinder angle and we include other coefficients in $g(\mathbf{r})$.
Equation~\ref{eq:twistwave} shows that the electron moves back-and-forth with both finite momentum ($\pm k$) and finite OAM ($\pm l$, $l$ is an integer) in a chiral box (\textbf{Figure~\ref{fig:orbital}b}), where $l$ changes sign if reversing $k$. Consequently, one can find that $k$ and $l$ are parallel (anti-parallel) locked to each other if we twist the box in a right(left)-handed way. The OAM corresponds to the self-rotation of the wavefunction around its center. We note that there are usually multiple $l$ values in Eq.~\ref{eq:twistwave}. Then, the total $l$ value is not necessarily a quantized number. 

We identify such orbital-momentum locking as the characteristic feature of the wavefunction within a chiral molecule. Similar to the spin-momentum locking in the Weyl fermion or CPL, we define the electronic chirality as right-handed or left-handed upon whether its orbital is parallel or anti-parallel to the momentum. Through the twisted box model, the structural chirality, i.e., the direction of twisting, determines the electronic chirality. We can define the electronic chirality by
\begin{equation}
    \chi_e = \mathrm{sign}(\mathbf{k \cdot l})
\end{equation}
where $\chi_e = +/- $ corresponds to right/left-handedness in electronic states.

\subsection{Orbital Texture in Chiral Materials}
We can observe the same orbital-momentum locking character in a realistic chiral molecule or crystal, for instance, from the wavefunction calculated by the state-of-art density-functional theory (DFT). It is tedious to separate two counter-propagating waves in a molecular wavefunction (see  Ref.~\cite{wan2023anomalous} for example). Instead, we take an infinite chain of the right-handed peptide $3_{10}$ helix as an example \cite{liu2021chirality}. It is a typical secondary structure found in proteins and polypeptides and was studied in some CISS experiments\cite{Tassinari2018}. Thanks to the well-defined lattice momentum by Bloch's theorem, we can easily obtain the Bloch wavefunction $\ket{u_k}$ at momentum $k$ and extract its orbital $L_z(k)=\bra{u_k} \hat{L}_z\ket{u_k}$ where $\hat{L}_z$ is the OAM operator. 
In a periodic chiral chain, $\ket{u_k}$ and $\ket{u_{-k}}$ are counter-propagating chiral waves. 
For a finite-size molecule, we can construct its wavefunction by superposing $\ket{u_k}$ and $\ket{u_{-k}}$ at discrete $k=pi/c$, similar to the twisted box model. 
In \textbf{Figures~\ref{fig:orbital}c} and \textbf{~\ref{fig:orbital}d}, one can find a general anti-symmetry among all bands, where $L_z$ flips in sign between $k$ and $-k$, i.e., the orbital-momentum locking. Additionally, $L_z$ changes the sign at given $k$ when we change the helix chirality,  which corresponds to reversing the electronic chirality.

Furthermore, the orbital-momentum locking can lead to a monopole-like orbital texture in the momentum space in a 3D chiral crystal \cite{yang2023monopole}, similar to the spin texture of a Weyl fermion. 
 \textbf{Figure~\ref{fig:orbital}d} shows the band structure and orbital texture of a chiral crystal PdGa. Despite the fact that PdGa exhibits a helix chain-like structure along [001] or [111] direction, its Fermi surface near the $\Gamma$ point exhibits an orbital-momentum texture along all directions. 

We point out that the existence of orbital texture unnecessarily requires a helical or spiral axis, despite the fact that a helical structure is vivid to illustrate rotating wavefunction. 
The orbital texture along an arbitrary direction away in PdGa also indicates this point. In general, the orbital-momentum locking, $\mathbf{l}(\mathbf{k}) = -\mathbf{l}(-\mathbf{k})$, is guaranteed by the time-reversal symmetry, which appears universally in nonmagnetic molecules or crystals. If the material is centrosymmetric, however, the inversion symmetry forces $\mathbf{l}(\mathbf{k}) = \mathbf{l}(\mathbf{-k})=0$. Thus, the minimal requirement of orbital-momentum locking is inversion-breaking (even not necessarily chiral) in a material. For instance, an achiral material that respects some mirror symmetry but breaks inversion can present orbital-momentum locking inside the mirror plane, \eg the van der Waals monolayer of MoS$_2$ or NbSe$_2$ \cite{xi2016ising}. In this case, the orbital is actually perpendicular to the momentum, and thus, the electronic chirality vanishes ($\chi_e = 0$). Orbital-momentum locking has less requirement in symmetry breaking than electronic chirality, which indeed needs a chiral material. 

Chirality requires breaking both mirror symmetry and inversion symmetry in the material. Chirality is a scalar quantity and remains unchanged for arbitrary rotation except for the inversion or mirror reflection. Therefore, the chirality presents a robust property for an ensemble of molecules (\eg gas, liquid, or a film)
where molecules orient randomly. Here, the vector quantity, such as the charge dipole, usually diminishes on average while the molecule chirality remains the same.  
For convenience, we summarize the symmetry properties of structural chirality ($\chi$), electronic chirality ($\chi_e$), and other related quantities in Table~\ref{tab:symmetry}, where $\chi_e$ behaves the same as $\chi$ upon different symmetry transforms.

\begin{table}[]
\centering
\begin{tabular}{c|cccc}
\hline
      Symmetry      &   $\mathcal{P}$ & $\mathcal{M}_x$ & $\mathcal{T}$ & $\mathcal{R}$ \\
            \hline
    $\chi$    &  $-\chi$ & $-\chi$ & $\chi$ & $\chi$\\
    $\chi_e =\mathrm{sign}(\mathbf{k\cdot l})$  &  $-\chi_e$ & $-\chi_e$ & $\chi_e$ & $\chi_e$\\
    $\mathbf{k}(k_x,k_y,k_z)$ & $(-k_x,-k_y,-k_z)$ & $(-k_x,k_y,k_z)$ & $(-k_x,-k_y,-k_z)$ & $\mathcal{R}\mathbf{k}$\\
    $\mathbf{l}(l_x,l_y,l_z)$ & $(l_x,l_y,l_z)$ & $(l_x,-l_y,-l_z)$ & $(-l_x,-l_y,-l_z)$ & $\mathcal{R}\mathbf{l}$ \\
    $\mathbf{s}(s_x,s_y,s_z)$ & $(s_x,s_y,s_z)$ & $(s_x,-s_y,-s_z)$ & $(-s_x,-s_y,-s_z)$ & $\mathcal{R}\mathbf{s}$\\
    \hline
\end{tabular}
\caption{The symmetry transform for the structural chirality($\chi$) and electronic chirality($\chi_e$) upon the inversion ($\mathcal{P}$), mirror reflection ($\mathcal{M}_x$), time-reversal ($\mathcal{T}$) and an ordinary rotation ($\mathcal{R}$). We show the $x\rightarrow -x$ reflection ($\mathcal{M}_x$) for the mirror without losing generality. 
We also show the transform of momentum ($\mathbf{k}$), orbital ($\mathbf{l}$) and spin ($\mathbf{s}$), where $\mathbf{l}$ and $\mathbf{s}$ transform in the same way. Upon $\mathcal{P}$ or $\mathcal{M}_x$, $\chi_e$ changes sign because $\mathbf{k}$ and $\mathbf{l}$ transform oppositely. In contrast, $\chi_e$ and $\chi$ remain the same upon $\mathcal{R}$ because $\mathbf{k}$ and $\mathbf{l}$ rotate in the same way.}
\label{tab:symmetry}
\end{table}

\subsection{Orbital Polarization Effect} \label{sec:OPE}

We reveal the orbital-momentum locking in the wavefunction as an equilibrium property. The net orbital polarization should be zero because opposite orbitals are degenerate due to time-reversal symmetry. Aiming to probe the electronic chirality, we need to push the system out of equilibrium, \eg by driving a current flow or irradiating light. When an electron transmits through a chiral molecule, it can inherit the electronic chirality and exhibit finite orbital polarization \cite{liu2021chirality}. We should acknowledge that a similar orbital polarization effect was proposed in a scenario of helical potential scattering by Nitzan \etal \cite{Gersten2013}. Compared to the scattering picture, orbital-momentum locking provides a way to understand transport and optoelectric effects from the intrinsic material properties, as we will discuss in the following.

\section{CHIRALITY INDUCED SPIN SELECTIVITY}

\begin{figure}
    \centering
    \includegraphics[width=0.8\textwidth]{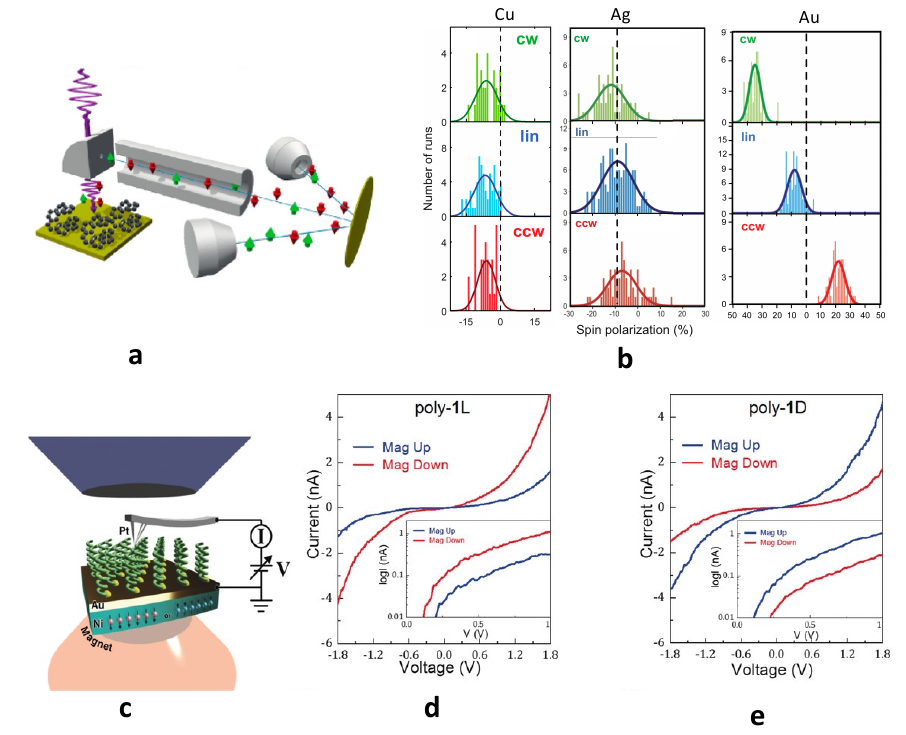}
    \caption{Experimental detection of CISS. (a) The photoemission setup with a Mott spectrometer probing the spin polarization. (b) Spin-polarized photoelectrons measured from different substrates (Cu, Ag, and Au) covered by M-hepthelicence. Here ``lin'' (cw or ccw) stands for linearly (clockwise or counter-clockwise circularly) polarized light.
    (c) The conductive atomic force microscopy (cAFM) setup. Except for cAFM, CISS was also measured in molecular spin-valve devices (see \textbf{Figure~\ref{fig:XiongExp}}). (d)-(e) The current-voltage relation measured upon switching the electrode magnetization for chiral polymers(poly-1L/1D). Each curve is averaged over about 100 measurements. The nearly zero-bias region was shown in the inset. Adopted from Refs.~\cite{Kettner2018,Mishra2020b}. }
    \label{fig:CISSexp}
\end{figure}

\subsection{Brief Overview}
The CISS effect was first observed in photoelectrons emitted from a gold substrate covered by a self-assembled monolayer of DNA\cite{Gohler2011}, in which the spin polarization was probed by the Mott polarizer. Since then, CISS has been revealed in a variety of chiral molecular systems\cite{Xie2011,Mishra2013,nino2014enantiospecific,Kettner2015,mondal2015field,Varade2018,Kettner2018,
Abendroth2019,Xiong2020,huizi2020ideal,mondal2021spin} via electrical, optical, or electrochemical probes. More broadly, CISS is shown to alert enantio-selective chemical and biological reactions ~\cite{naaman2020chiral,ghosh2022control} and facilitate enantiomer separation ~\cite{BanerjeeGhosh2018,safari2023enantioselective} and unconventional superconductivity ~\cite{Alpern2016,shapira2018unconventional,wan2023signatures}. 
Most of these experiments were performed at room temperature, suggesting a highly consequential interaction between molecular structural chirality and electron spin.

In transport experiments, CISS is commonly detected by the two-terminal  magnetoresistance (MR) \cite{Xie2011,Kiran2016,Varade2018,Xiong2020,Mishra2020b,
Kim2021,Qian2022} where the chiral molecule links a ferromagnetic substrate and a nonmagnetic lead (see \textbf{Figure~\ref{fig:CISSexp}}). The resistance changes as flipping the substrate magnetization ($M$). Because the MR is similar to that of a spin valve, chiral molecules were usually presumed to generate spin polarization in transport. The CISS MR is found to increase with molecule length and can be much beyond 60\% in reports~\cite{Kulkarni2020,Mishra2020,Al-Bustami2022}, although the ferromagnetic electrode exhibits much smaller spin polarization at the Fermi energy, \eg about 20\% for Ni. 

Despite increasing experimental results and theoretical efforts, the microscopic origin and physical mechanisms behind CISS remain open questions thus far~\cite{evers2021theory}. It is clear that chiral materials such as DNA exhibit no spin polarization at equilibrium, and CISS refers to generating spin polarization in the nonequilibrium process. In principle, spin-orbit coupling (SOC) is required to couple the structural chirality (pathway of electron motion) and the electron spin. However, the SOC of organic materials composed of light elements is negligible compared to the room temperature (26 meV) and thus is too weak to account for the room-temperature robust spin selectivity. 
Therefore, a number of theoretical approaches were proposed to enhance the SOC within chiral molecules, primarily on spin-dependent scattering and tight-binding models 
\cite{Yeganeh2009,Guo2012,Gutierrez2012,Medina:2012ud,Guo2014,Medina2015,Matityahu2016,Dalum2019,Michaeli2019,Dubi2021} by introducing dephasing/dissipation
\cite{Guo2014,Matityahu2016,Diaz2018b,volosniev2021interplay}, electron-electron correlation~\cite{fransson2019chirality,Fransson2021},  electron-vibration/polaron interaction ~\cite{du2020vibration,zhang2020chiral,wu2021electronic,das2022temperature,Klein2023}, etc. 

An alternative approach requires no SOC from chiral molecules. Nitzan \etal\cite{Gersten2013} proposed that electron OAM gets polarized after being scattered by the chiral potential, and the orbital polarization induces spin selectivity due to the strong SOC from the metal substrate. This proposal, however, was questioned because CISS was observed in photoemission experiments where substrates (\eg Al or Cu) have negligible SOC~\cite{Mishra2013,Kettner2018}. Later, Liu \etal \cite{liu2021chirality} pointed out an important difference between photoemission
and transport detection for CISS (\textbf{Figure~\ref{fig:CISSexp}}). The former detects the global angular momentum that includes both orbital and spin because photoelectrons in the vacuum can carry quantized angular momentum known in earlier theory\cite{Bliokh2007} and experiments \cite{Uchida2010,lloyd2017}. In contrast, magneto-transport probes the spin exclusively via the ferromagnetic electrode. Thus, Liu \etal proposed that electrodes with varied SOC should be investigated to examine this mechanism because no control measurement on electrodes had been done in CISS transport by that time~\cite{liu2021chirality}. 
In addition, other models related to the electrode SOC ~\cite{Dubi2021,Hegard2023} were also proposed recently.

\begin{figure}
    \centering
    \includegraphics[width=1\textwidth]{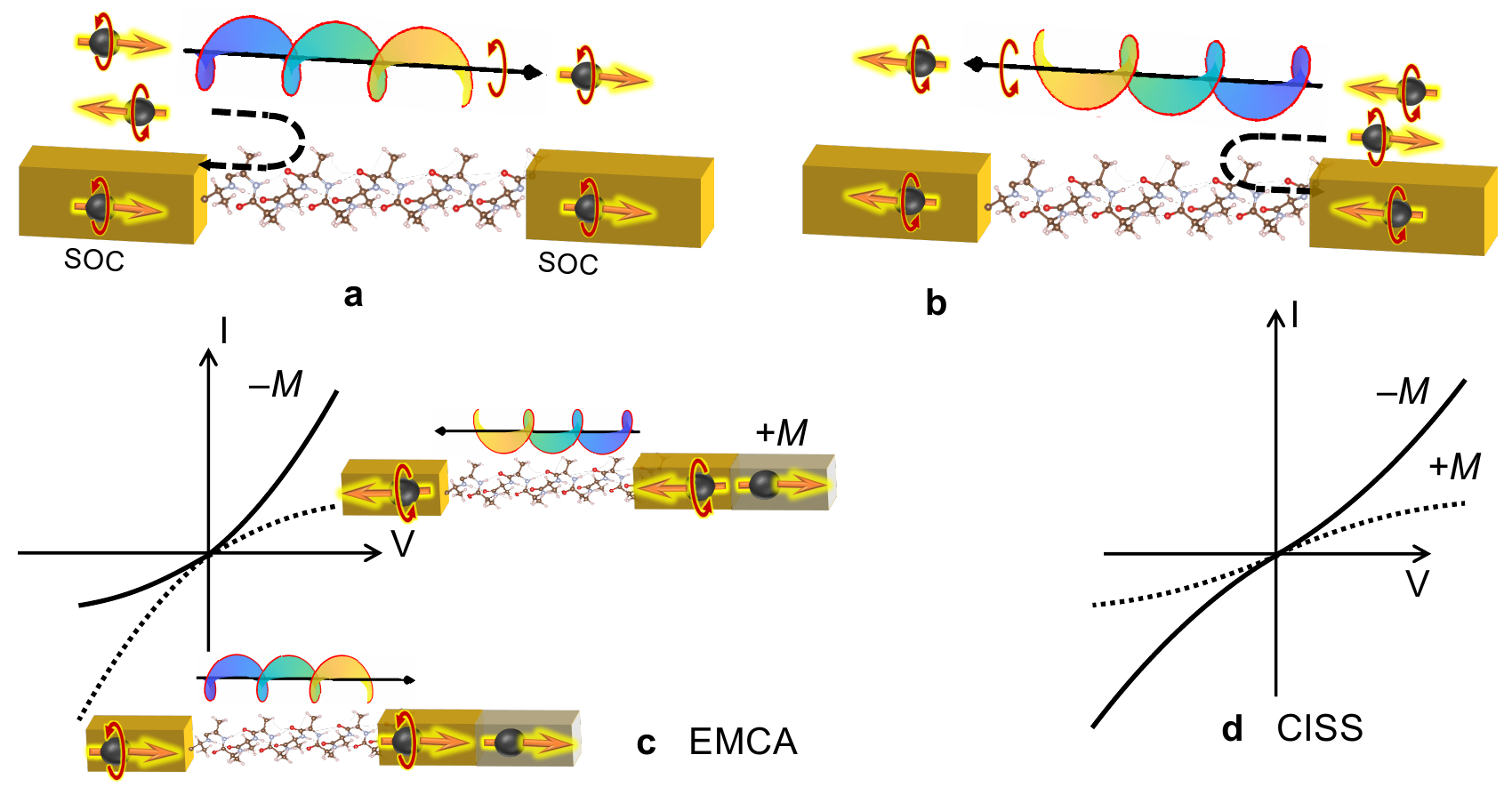}
    \caption{Schematics of the orbital and spin polarization in transport. (a)-(b) The electrode SOC locks spin and orbital together, and the chiral molecule alters the orbital. From the incident electrode, the chiral molecule filters the orbital and thus also filters the spin. In the outgoing electrode, SOC generates spin polarization from the orbital polarization induced by the chiral molecule. We note that reflected electrons exhibit the same spin polarization as transmitted ones due to a SOC-driven spin flip upon reflection. (c) Schematics of the I-V relation for EMCA. For instance, the positive (negative) bias leads to negative (positive) spin polarization [shown in (b)/(a)] and induces low (high) conductance when attaching to a ferromagnetic contact with $+M$ polarization. (d)  Schematics of the I-V relation for CISS (also see \textbf{Figure~\ref{fig:CISSexp}d/e}), which is different from EMCA and cannot be naively explained by the spin filter/polarizer regardless of SOC origin. 
    }
    \label{fig:orbitalpolarizer}
\end{figure}

In addition, a spin filter scenario was widely presumed regarding CISS~\cite{Naaman2012,Naaman2019}, in which the transmitted and reflected electrons exhibit opposite spin polarization, was found to generate unphysical consequences in recent reports \cite{Yang2019,Yang2019b,Wolf2022}.  
Instead, a spin polarizer picture was suggested where the transmitted and reflected electrons should exhibit the same spin polarization in CISS, as illustrated in \textbf{Figures~\ref{fig:orbitalpolarizer}a} and \textbf{~\ref{fig:orbitalpolarizer}b}. It indicates that the CISS effect can induce spin polarizations in two electrodes with the same sign provided two electrodes have similar SOC~\cite{Wolf2022}. 

We summarize two major questions to understand CISS. 
\begin{itemize}
    \item What is the origin of the strong SOC (compared to room temperature)? 
    \item Why is the CISS spin polarization characterized by MR so large (even larger than the ferromagnet contact)? 
\end{itemize}
Even if the first question is answered to realize a robust spin polarizer,  the second question alone is still nontrivial~\cite{xiao2022nonreciprocal}. In a perturbative scenario, a spin filter/polarizer was found to generate EMCA with a low MR (several percent or less) in a chiral conductor \cite{liu2021chirality,Yang2019b}, as illustrated in \textbf{Figure~\ref{fig:orbitalpolarizer}c}. However, 
the CISS MR displays a fundamentally distinct symmetry (see \textbf{Figure~\ref{fig:orbitalpolarizer}d} or \textbf{\ref{fig:CISSexp}b/c}) by circumventing the constrain of Onsager's reciprocal relation\cite{Dalum2019,Yang2019b,Yang2019,Naaman2020comment,Yang2020reply,xiao2022nonreciprocal}, yet to be understood.

\subsection{Spin, Orbital, and Spin-Orbit Coupling in CISS}
The orbital polarization effect due to the orbital-momentum locking suggests that chiral molecules act as an orbital polarizer rather than a spin polarizer/filter,
and the electrode SOC converts the orbital polarization into spin polarization, thus producing CISS without necessarily requiring SOC from molecules, as illustrated by \textbf{Figure~\ref{fig:orbitalpolarizer}a/b}. 
In other words, the chiral molecule together with electrodes can be viewed as a spin polarizer.

\begin{figure}
    \centering
    \includegraphics[width=\textwidth]{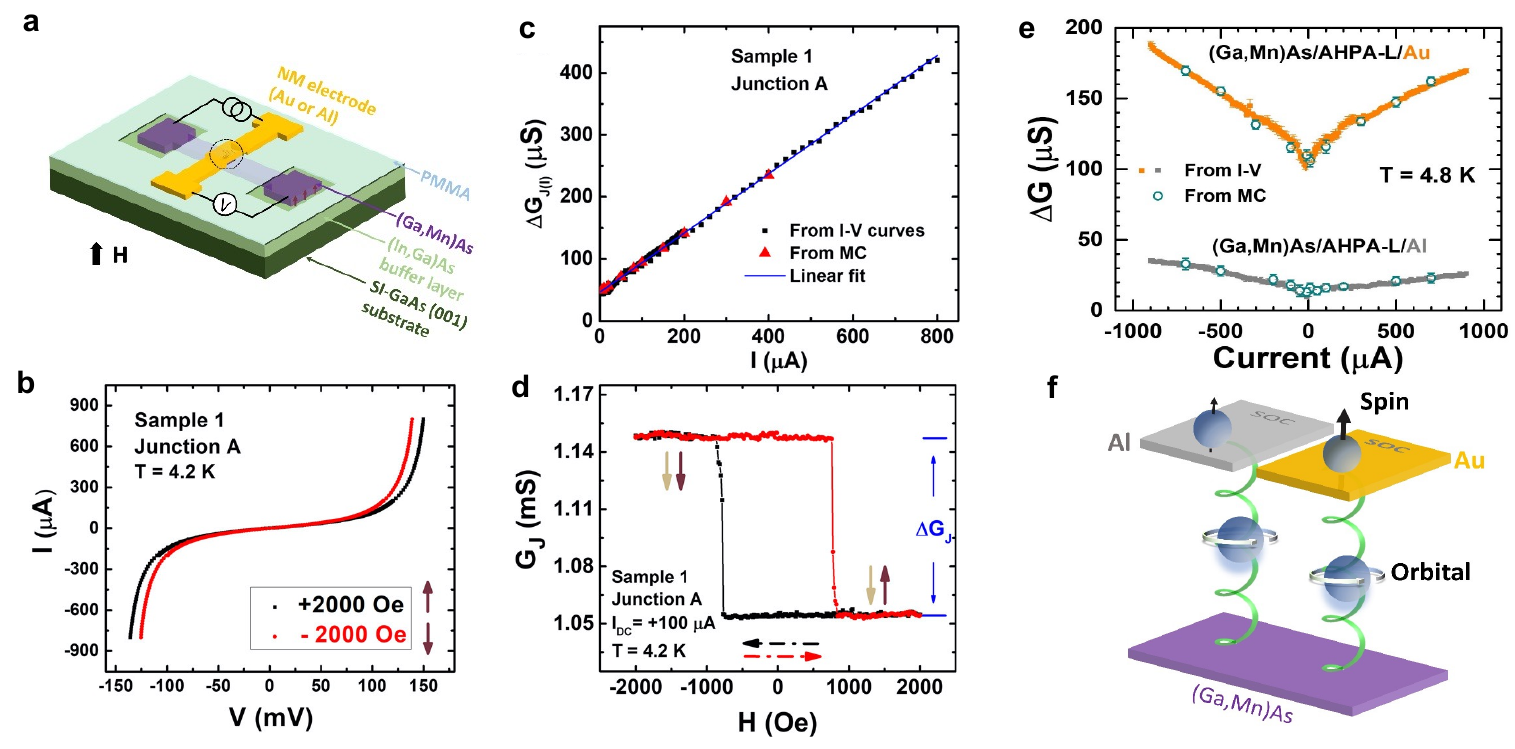}
    \caption{The spin valve device with chiral molecular junctions. (a) Schematic diagram of the device structure along with the junction measurement setup. (b) I-V characteristics of the junction in perpendicular magnetic fields. (c) The dependence of magnetoconductance ($\Delta G$), \ie the conductance change between opposite magnetic fields, on the current. $\Delta G$ is nonzero at the zero current/bias limit, deviating from Onsager's reciprocal relation.  (d) The Junction conductance versus perpendicular magnetic field was measured at a DC bias of 100 $\mu$A. The $\Delta G$ is marked in the hysteresis and represents the CISS effect. (b)-(d) correspond to devices with the Au electrode. (e) Compare $\Delta G$ between devices with Au and Al electrodes. The Al electrode case exhibited a largely suppressed CISS effect.   (f) Schematic of the orbital to spin conversion by electrode SOC where chiral molecules polarize the orbital. Adopted from Refs.~\cite{Xiong2020,Xiong2023interplay}.
    }
    \label{fig:XiongExp}
\end{figure}

Most CISS spin valve devices employed electrodes involving heavy metals (\eg Au or Pt). For example, the conductive atomic force microscopy (cAFM) setup~\cite{Xie2011} has been frequently used to demonstrate CISS (see \textbf{Figure~\ref{fig:CISSexp}a}). The cAFM tip, which is usually made of Pt, acts as the nonmagnetic electrode, and a magnetic substrate (such as the Ni film covered by a thin protection layer of Au) acts as the ferromagnetic electrode. Although its implementation is relatively straightforward, cAFM experiment relies on averaging a large number of current-voltage (I-V) measurements to mitigate the fluctuation and instability (see \textbf{Figure~\ref{fig:CISSexp}b/c} for example). In contrast, thin film-based molecular junctions, in which a chiral layer is sandwiched between two conducting electrodes, are more stable for reproducible I-V curves. 
Xiong group and co-workers recently demonstrated such molecular junctions by using a ferromagnetic semiconductor (Ga, Mn)As with preferred out-of-plane magnetization as the magnetic spin analyzer and another Au electrode. They presented a first rigorous determination of the I-V relation due to CISS~\cite{Xiong2020} (\textbf{Figure~\ref{fig:XiongExp}}). 

Motivated by the orbital polarization effect, Xiong group and collaborators \cite{Xiong2023interplay}
compared different nonmagnetic electrodes in their molecular spin valves. The experiments yielded quantitative differentiation of the magnitude and bias-dependence of the spin valve conductance in junctions with electrodes of contrastingly different SOC strengths (Au versus Al) and layers of chiral and achiral molecules. Their results revealed a definitive correlation between the magnitude of the CISS spin valve conductance and the SOC strength in the nonmagnetic electrode: The molecular junctions with Au electrodes exhibit significant MC whose magnitudes depend distinctly on the chirality or length of the molecules; in contrast, in otherwise identical devices with Al electrodes, regardless of the molecule involved the MC are essentially indistinguishable from those of the control samples without any molecules, as shown in \textbf{Figure~\ref{fig:XiongExp}e-f}. The work unambiguously evidenced the essential role of the contact SOC in generating observable CISS effect in chiral molecular spin valves. 

Furthermore, the CISS effect was measured in molecular junctions below 5 K~\cite{Xiong2020,Xiong2023interplay} and also in single molecules by a high-quality spin-polarized scanning tunneling microscopy at 5 K~\cite{Safari2023spin}. The existence of CISS MR at such low temperatures excludes the electron-phonon/vibration interaction from primary CISS mechanisms in these experiments. The single molecule measurement also excludes the ensemble effect (inter-molecule interaction) here.

After chiral molecules polarize the orbital, the electrode SOC translates the orbital polarization to spin polarization in the molecular junction. This idea even inspired a strategy to detect the orbital current generated from the long-sought orbital Hall effect in ordinary metals or semiconductors \cite{Bernevig2005,Tanaka2008}. A thin layer of heavy-metal such as Pt with strong SOC was proposed to cover the orbital Hall effect film (\eg Cu or Al) and convert the transverse orbital current to spin current at the interface \cite{xiao2022detection}, which was realized in a recent spintronic experiment~\cite{Rothschild2022}. 

Additionally, the phonon OAM ~\cite{Zhang2015phonon,Chen2019phonon,Grissonnanche2020phonon,Chen2021phonon} in inversion-breaking materials, including chiral materials, has garnered growing attention too. Recently, a spin Seebeck effect~\cite{Kim2023phonon} was observed where electron spin polarization is generated from the phonon orbital polarization in the presence of the temperature gradient.

\section{EMCA vs CISS}

\subsection{Electric MagnetoChiral Anisotropy } \label{sec:emca}
In transport, another known chirality-induced phenomenon is EMCA\cite{Rikken2001}. It refers to the resistance that depends on the current ($\mathbf{I}$) and magnetic field ($\mathbf{B}$) for a chiral conductor, 
\begin{equation} \label{eq:emca}
 R^{\chi}=R_0(1+\alpha B^2 + \beta^{\chi} \mathbf{B \cdot I})   
\end{equation}
where $\beta^{\chi}=-\beta^{-\chi}$ and
{$\chi = \pm$ stands for chirality}. The $B^2$ term represents the ordinary MR while the $\mathbf{B \cdot I}$ term represents EMCA with unidirectional resistance like a diode. The EMCA was observed in many chiral crystals, organic chiral conductors, and even carbon nanotubes
~\cite{Rikken2001,krstic2002magneto,pop2014electrical,Rikken2019,Inui2020,Shiota2021,Ye2022Te}, where the MR is usually a few percent or even less. We stress that the material magnetization $\mathbf{M}$ can also replace the $\mathbf{B}$ field in Eq.~\ref{eq:emca}, as another way to break the time-reversal symmetry \cite{pop2014electrical,Aoki2019,Yokouchi2017}. 

\begin{figure}
    \centering
    \includegraphics[width=0.6\textwidth]{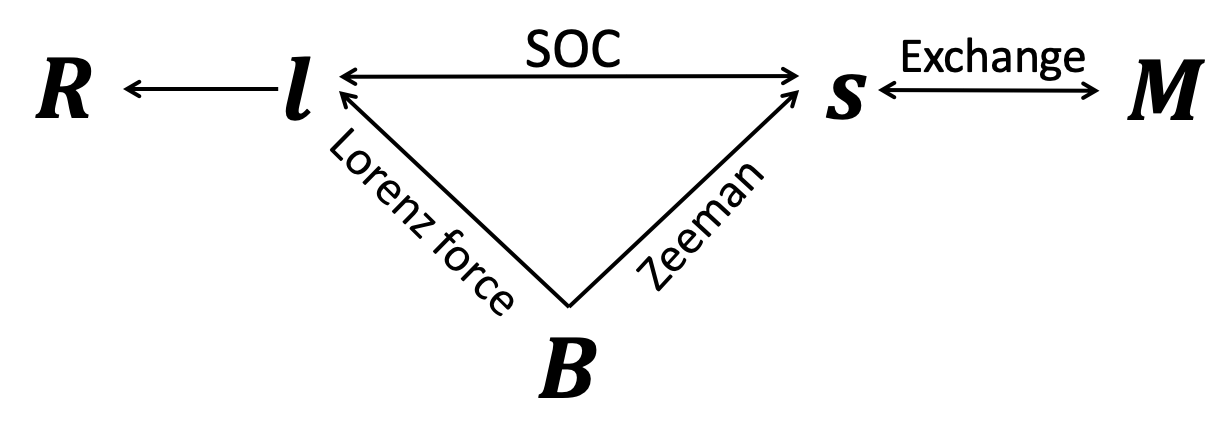}
    \caption{Interactions among the magnetic field ($\boldsymbol{B}$), electrode magnetization ($\boldsymbol{M}$), orbital ($\boldsymbol{l}$), spin ($\boldsymbol{s}$) and resistance ($\boldsymbol{R}$). Here, $\boldsymbol{l}$ alters the charge transport, \ie $\boldsymbol{R}$. $\boldsymbol{B}$ couples to $\boldsymbol{l}$ directly while $\boldsymbol{M}$ interacts with $\boldsymbol{l}$ via $\boldsymbol{s}$ assisted by the spin-orbit coupling (SOC). In CISS transport, $\boldsymbol{R}$ rather than $\boldsymbol{s}$ is usually directly measured.
    }
    \label{fig:coupling}
\end{figure}

EMCA is also called nonreciprocal MR in condensed matter physics literature\cite{Tokura2018review}. Theories based on the semiclassical transport~\cite{Ideue2017,Rikken2019} and quantum geometry~\cite{kaplan2022unification} were developed 
to explain EMCA in solids. 
Alternatively, the orbital-momentum locking provides an illustrative picture for EMCA. As discussed in Sec.\ref{sec:OPE},
the current drives electron motion (momentum) and generates orbital polarization. We note that $\mathbf{B}$ couples directly to the electron orbital by the Lorenz force while $\mathbf{M}$ couples to the orbital via SOC, eventually leading to the resistance change, as shown in \textbf{Figure~\ref{fig:coupling}}.
In the case of $\mathbf{B}$, we expect low (high) resistance if the Lorenz force-induced orbital aligns in the same (opposite) direction to the current-induced orbital. Therefore, the total resistance depends on $\mathbf{B\cdot I}$. 
In the case of $\mathbf{M}$, we recall that the orbital polarization effect can be converted to spin polarization by SOC, in which $\pm \mathbf{I}$ give rise to opposite spin polarization (see \textbf{Figures~\ref{fig:orbitalpolarizer}a} and \textbf{~\ref{fig:orbitalpolarizer}b}).  Here,  $\mathbf{M}$ or SOC may come from either the electrode or chiral conductor. Then, the resistance depends on whether the current-induced spin polarization is parallel or anti-parallel to $\mathbf{M}$, as demonstrated in \textbf{Figure~\ref{fig:orbitalpolarizer}c}. Quantum transport calculations~\cite{Yang2019b,liu2021chirality} indeed showed that the CISS spin polarizer/filter with a magnetic electrode presents the EMCA behavior. 

Additionally, EMCA can be generalized to achiral but noncentrosymmetric materials such as surfaces/interfaces\cite{Rikken2005,choe2019gate} and polar conductors\cite{Ideue2017} with an extra term $\mathbf{B\times I}$ in Eq.~\ref{eq:emca}. This is consistent with the discussion (Sec.~\ref{sec:locking}) that the orbital-momentum locking appears in general inversion-breaking systems where the orbital and momentum can sometimes be perpendicular to each other.

\subsection{Onsager's Reciprocal Relation}

We highlight that EMCA and CISS MR are fundamentally different transport phenomena in symmetry. EMCA respects Onsager's reciprocal relation, while CISS MR violates such reciprocity. \textbf{Figure~\ref{fig:orbitalpolarizer}} illustrates the I-V relation of CISS MR and EMCA. Onsager's reciprocal theorem originates in the microscopic reversibility of thermodynamic equilibrium and poses strict constraints on macroscopic conductivity. In the small bias limit, the reciprocity forces that two-terminal conductance remains unchanged as reversing time, i.e., $G(\mathbf{I,B/M})=G(\mathbf{-I,-{B}/-{M}})$ \cite{Onsager1931,LANDAU1980}, which holds for EMCA (see Eq.~\ref{eq:emca}) but not for CISS MR~\cite{Rikken2023}. Therefore, EMCA can be derived by weakly perturbing the equilibrium ground state in the semiclassical theory~\cite{Ideue2017,Liu2021UMR,kaplan2022unification} or Landauer-B\"utikker transport calculations~\cite{liu2021chirality}.
However, CISS MR may require understanding the far out-of-equilibrium phase. 

CISS MR experiments ~\cite{Kiran2016,Naaman2020comment,Xiong2020,Xiong2023interplay} witnessed a clear deviation from the Onsager's relation by showing the zero-bias magneto-conductance, $G(\mathbf{M})\neq G(-\mathbf{M})|_{V/I\rightarrow0}$, \eg see \textbf{Figures~\ref{fig:CISSexp}b/c} and \textbf{~\ref{fig:XiongExp}c/e}. It indicates that switching $\mathbf{M}$ drives the system to different nonequilibrium states, consequently leading to varied conductance. If so, what kind of nonequilibrium states matter here? Given that a CISS device and EMCA share the same symmetry condition, both inversion/mirror symmetry-breaking and time-reversal symmetry-breaking, what else reasons cause different transport behaviors between CISS and EMCA? Answers to these questions may help us understand the nature of CISS. In experiments, EMCA involves metals or doped semiconductors, while CISS MR refers to insulating chiral molecules (\eg the typical resistance is in the order of magnitude of $\mathrm{G\Omega}$ in CISS as shown in \textbf{Figure~\ref{fig:CISSexp}b/c}). Thus, the insulating nature of chiral molecules may be essential for understanding CISS MR, which has rarely been appreciated thus far~\cite{xiao2022nonreciprocal}. 

In CISS MR, the spin polarizer/filter (with the rationalized source of SOC) cannot solely resolve the conflict with the Onsager's relation. As discussed in Sec.\ref{sec:emca}, the spin polarization in the context of CISS directly leads to the EMCA transport, consistent with recent experiments on chiral conductors. To understand the CISS MR, however, we need a mechanism beyond the spin polarizer/filter because of the fundamental symmetry reason. Along this line, some models were recently proposed, \eg the spinterface\cite{Dubi2021,alwan2023temperature}, 
spin-dependent barrier~\cite{das2022temperature},
interface spin-transfer-torque~\cite{Hegard2023}, and charge-trapping-induced tunneling barrier modulation~\cite{xiao2022nonreciprocal}. Beyond rationalizing Onsager's relation, a successful theory should also clarify why the chiral conductor exhibits EMCA while the chiral molecule shows CISS-type MR~\cite{xiao2022nonreciprocal}. 

\subsection{Spin Polarization in CISS transport}
The spin polarization ($P$) was commonly adopted as a figure of merit to characterize the amplitude of CISS. Nevertheless, most CISS experiments probe the spin polarization in an indirect way. In transport, 
the normalized MR, 
\begin{equation}
    \widetilde{MR}=|\frac{R(-M)-R(+M)}{R(-M)+R(+M)}|\approx|\frac{I(-M)-I(+M)}{I(-M)+I(+M)}|
\end{equation}
 was commonly considered to be equivalent to the spin polarization ratio ($P$) induced by the chirality, similar to the earlier description of the tunneling MR \cite{Julliere1975TMR}. However, $P \equiv \widetilde{MR}$ assumes 100\% polarization in the ferromagnetic electrode, which is apparently unrealistic. 
 
It has been well known that MR can be much larger than the electrode spin polarization in the MgO-based tunneling MR~\cite{Butler2001,Mathon2001}, a milestone in spintronics. 
So, we cannot naively presume $P \equiv \widetilde{MR}$ as a ubiquitous rule to interpret the MR. 
In the tunneling MR,, the nonlinearity of the barrier tunneling significantly enhances MR so that $\widetilde{MR} \gg P$, which motivated the tunneling barrier modulation model~\cite{xiao2022nonreciprocal} to understand the CISS MR recently.

Furthermore, the spin becomes a non-conserved quantity in a CISS device with significant SOC because SOC mixes spin up and down channels. In this case, it becomes ambiguous to argue independent spin channels when rationalizing experiments that actually measure the transport of charge rather than spin. 

Overall, the spin polarization scenario deserves more caution than it does to thoroughly understand CISS experiments \cite{Liu2023spin}. 
Instead, studies of the CISS mechanism should aim to understand the directly observed quantities, \eg the charge transport rather than spin polarization (also see \textbf{Figure~\ref{fig:coupling}}).

\section{ANOMALOUS CIRCULARLY POLARIZED LIGHT EMISSION}

\begin{figure}
    \centering
    \includegraphics[width=1.\textwidth]{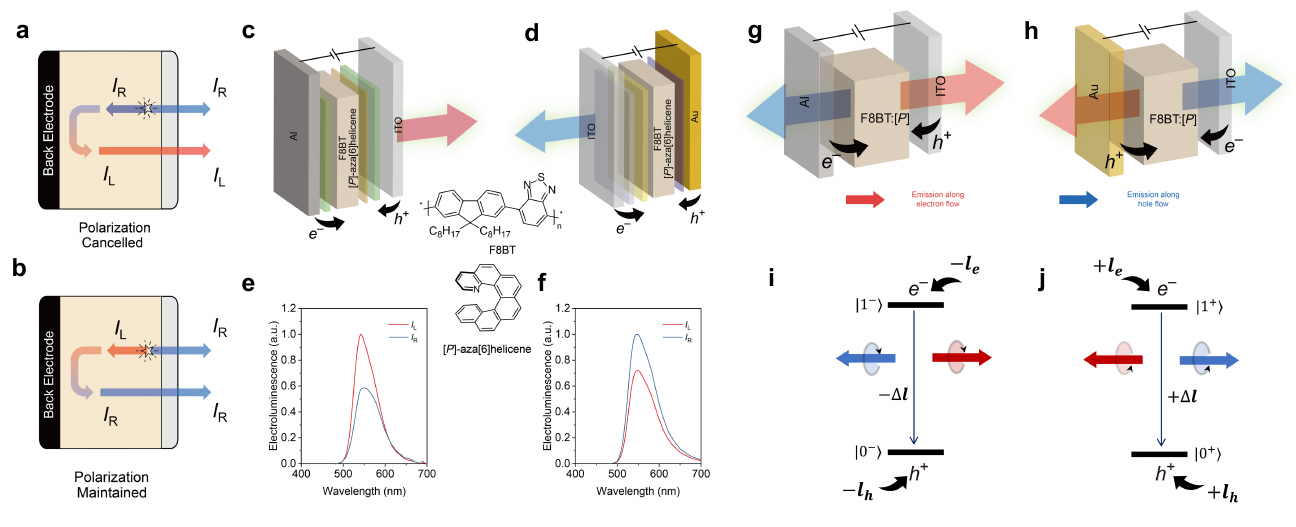}
    \caption{ Circularly polarized light (CPL) emission from the chiral organic light-emitting diode (OLED).
    (a) Normal CPL emission where the handedness is independent of the emission direction, where $I_L$ and $I_R$ represent the light intensity of the left-handed (red arrows) and right-handed (blue arrows) CPL. (b) Anomalous CPL emission where the handedness is dependent on the emission direction. The star-shaped symbol indicates the emission sites in devices. (c) Conventional and (d) inverted device structures for OLEDs, where the insets show the molecular structures in chiral polymers. The CPL emission was recorded from conventional (e) and inverted (f) OLEDs. (g)-(h) Simultaneous CPL emission from both sides of the OLED where the CPL handedness depends on the emission direction and current-flow direction. (i)-(j) The electron-hole recombination mechanism, including the angular momentum transfer, to explain the anomalous CPL emission. The electron ($e^-$)/hole ($h^+$) carries momentum-dependent orbital ($\pm l_e / \pm l_h$) as current flows due to the orbital-momentum locking in the chiral material. Photons spontaneously emitted in opposite directions carry the same angular momentum ($\pm \Delta l$), thus exhibiting opposite handedness.
    }
    \label{fig:CPL}
\end{figure}

It is long known that chirality interacts with CPL intimately. The chiral media selectively adsorbs the CPL with opposite handedness, called the circular dichroism, which serves as a convenient way to measure the structural chirality. As an inverse effect of circular dichroism, the chiral material can illuminate CPL in circularly polarized photoluminescence (CP-PL) or electroluminescence (CP-EL). In photoluminescence, the CPL handedness is solely determined by the material chirality and presumed to be irrelevant to the luminescence direction. In the textbook, circular dichroism and CP-PL/EL are well explained by the product of the electric and magnetic transition dipole moments~\cite{craig1984molecular}.

Surprisingly, the electronic chirality, \ie orbital-momentum locking, leads to a new chiroptical interaction called the anomalous CPL emission (ACPLE) \cite{wan2023anomalous}. Distinct from ordinary CP-PL/EL, ACPLE shows that the CPL handedness can be further controlled by the current flow or light-emission direction, besides the chirality. It is closely related to the Berry phase, a topological electronic property, in the optical transition. 

Wan \etal fabricated organic light-emitting diodes (OLEDs) based on chiral polymers~\cite{Wan2020,wan2019inverting,wan2021strongly,Lee2017a,DiNuzzo2017} and discovered the ACPLE \cite{wan2023anomalous}. The CP-EL exhibits opposite handedness in forward and backward emission directions, counter-intuitive to what is usually expected in EL or PL (see \textbf{Figure~\ref{fig:CPL}g/h}). With such direction-dependent CPL emission, the back-reflected light exhibits the same handedness as the forward emission, avoiding the polarization cancellation that occurs in OLEDs using other materials and boosting the net CPL polarization~\cite{yan2019configurationally,zinna2017design}. 
Furthermore, the current flow can also switch CPL handedness in an OLED. 

The ACPLE can be understood via the orbital-momentum locking in chiral materials. 
The current flow induces nonequilibrium orbital polarization in electron and hole carriers. 
Therefore, finite angular momentum transfers from the electron/hole orbital to the photon spin.  In the optical transition with fixed angular momentum transfer, the counter-propagating photons carry the same spin, thus displaying opposite handedness by definition. Additionally, reversing the current flow changes the sign of angular momentum transfer, switching the CPL handedness, too. 

ACPLE reveals an exotic CP-EL mechanism caused by current-induced time-reversal breaking. The transfer angular momentum corresponds to the optical Berry phase in the electron-hole recombination. Liu \etal \cite{liu2023anomalous} recently generalized ACPLE to solids and re-formulated it by the optical Berry curvature dipole, an intrinsic property from the band structure, in noncentrosymmetric (including chiral) materials (\eg Refs.~\cite{Zhang2014,Pu2021WS2_exp})  
The ACPLE paves the way to design novel chiroptoelectronic devices with strong circular polarization. 
Furthermore, ACPLE can also be regarded as an inverse effect of the circular photogalvanic effect (CPGE), in which irradiation of CPL generates a dc current as a nonlinear light-matter interaction~\cite{Sipe2000}. 
In CPGE, the photon spin is transferred to the OAM of electrons and holes. 
Because finite OAM leads to finite velocity again due to orbital-momentum locking, 
electrons and holes can generate a DC current. 

In addition, Kim \etal \cite{Kim2021} designed spin light-emitting diodes based on the chiral metal-halide perovskite hybrid semiconductor. Here, CISS was believed to produce spin-polarized carriers, where organic molecules provide chirality and perovskites bring strong SOC. Because the current flow 
controls the spin polarization direction in CISS, we anticipate that the CPL handedness also relies on the current direction and light emission direction in these devices.

\section{OUTLOOK}
Chirality-driven phenomena, including CISS, EMCA, and ACPLE, demonstrate the intriguing interplay between different degrees of freedom, \ie chirality, spin, orbital, and charge, and carry profound and broad implications in quantum science and technology~\cite{aiello2022chirality,chiesa2023chirality}. 
Beyond organic molecules and polymers, 
hybrid organic-inorganic materials \cite{lu2019spin,Kim2021,Qian2022,Al-Bustami2022}, twisted van der Waals layers \cite{Liu2021UMR,zhang2022diodic} and chiral crystals~\cite{pop2014electrical,Rikken2019,Inui2020,Shiota2021,Ye2022Te,yang2023monopole,Kim2023} attract increasing interest recently. 

Moreover, the CISS effect paves novel pathways to control chemical and biological processes by the magnetic field, magnetism or CPL ~\cite{Banerjee2020,Naaman2022,Ghosh2022,li2023observation,zuo2023mechano}, such as enhancing the antibody-antigen association by a magnetic surface~\cite{Banerjee2020}, manipulating the protein activity by CPL~\cite{Ghosh2022}, and even providing a clue to understanding the origin of biomolecular homochirality in life\cite{ozturk2023origin}. 

Nevertheless, more efforts are required to establish a unified quantum theory framework in which the orbital effect may serve as a basic microscopic scenario to reveal the physics behind these exciting phenomena and predict new experiments for verification.

\section*{DISCLOSURE STATEMENT}
The authors are not aware of any affiliations, memberships, funding, or financial holdings that might be perceived as affecting the objectivity of this review. 

% Acknowledgements
\section*{ACKNOWLEDGMENTS}
We are grateful for fruitful collaborations and/or inspiring discussions with Tianhan Liu,
Yizhou Liu, Jiewen Xiao, Jahyun Koo, Yotam Wolf, Qun Yang, Yongkang Li, Noejung Park, Yuwaraj Adhikari, Peng Xiong, Jianhua Zhao, Li Wan, Matthew J. Fuchter, Amir Capua, Yuval Oreg, Ady Stern, Zhong Wang, Per Hedegard, Helmut Zacharias,  Yossi Paltiel,  and Ron Naaman. We acknowledge the financial support by the European Research Council (ERC Consolidator Grant ``NonlinearTopo'', No. 815869), the ISF - Personal Research Grant	(No. 2932/21), and the Minerva Foundation supported by BMBF.

% \\
%\bibliography{chiral-references}

\end{document}